# COMPARATIVE PERFORMANCE ANALYSIS OF DIFFERENT MODULATION TECHNIQUES FOR PAPR REDUCTION OF OFDM SIGNAL

Md. Munjure Mowla[1], Liton Chandra Paul[2] and Md. Rabiul Hasan[3]

[1,2,3] Department of Electronics & Telecommunication Engineering,
Rajshahi University of Engineering & Technology, Rajshahi, Bangladesh

## ABSTRACT

*One of the most important multi-carrier transmission techniques used in the latest wireless communication arena is known as Orthogonal Frequency Division Multiplexing (OFDM). It has several characteristics such as providing greater immunity to multipath fading & impulse noise, eliminating Inter Symbol Interference (ISI) & Inter Carrier Interference (ICI) using a guard interval known as Cyclic Prefix (CP). A regular difficulty of OFDM signal is high peak to average power ratio (PAPR) which is defined as the ratio of the peak power to the average power of OFDM Signal. An improved design of amplitude clipping & filtering technique of us previously reduced significant amount of PAPR with slightly increase bit error rate (BER) compare to an existing method in case of Quadrature Phase Shift Keying (QPSK) & Quadrature Amplitude Modulation (QAM). This paper investigates a comparative performance analysis of the different higher order modulation techniques on that design.*

## KEYWORDS

*Bit Error rate (BER), Complementary Cumulative Distribution Function (CCDF), Long Term Evolution (LTE), Orthogonal Frequency Division Multiplexing (OFDM) and Peak to Average Power Ratio (PAPR).*

## 1. INTRODUCTION

The quick growth in multimedia controlled applications has triggered an insatiable thirst for high data rates and resulted in an increased demand for technologies that support very high speed transmission rates, mobility and efficiently utilize the available spectrum & network resources. OFDM is one of the paramount resolutions to achieve this goal and it offers a promising choice for future high speed data rate systems [1].OFDM has been standardized as part of the IEEE802.11a and IEEE 802.11g for high bit rate data transmission over wireless LANs [2]. It is incorporated in other applications and standards such as digital audio broadcasting (DAB), digital video broadcasting (DVB), European HIPERLAN/2 and the Japanese multimedia mobile access communications (MMAC). In addition, OFDM is also used now as the transmission scheme of choice in the physical layer of the world wide interoperability for microwave access (WiMAX) & long term evolution (LTE) standards. It has also been used by a variety of commercial applications such as digital subscriber line (DSL), digital video broadcast- handheld (DVB-H) and Media FLO[3]. As the data rates and mobility supported by the OFDM system raise, the number of subcarriers also raise, which in turn leads to high PAPR. As future OFDM-based systems may push the number of subcarriers up to meet the higher data rates and mobility demands, there is a need to mitigate the high PAPR.

A number of attractive approaches have been proposed & implemented to reduce PAPR with the expense of increase transmit signal power, bit error rate (BER), computational complexity and data rate loss etc. So, a system trade-off is required. These reduction techniques are basically divided into three types of classes such as signal distortion, multiple signaling & probabilistic and coding. In this paper, amplitude clipping & filtering based design (signal distortion) is used to reduce PAPR with a little compromise of BER. The main objective of this paper is to investigate the comparative performance analysis of different higher order modulation technique on that particular design.

## 2. BASIC MODEL OF OFDM SYSTEM

OFDM is a special form of multicarrier modulation (MCM) with densely spaced subcarriers with overlapping spectra, thus allowing multiple-access. MCM works on the criteria of transmitting data by dividing the stream into several bit streams, each of which has a much lower bit rate and by using these sub-streams to modulate several carriers.

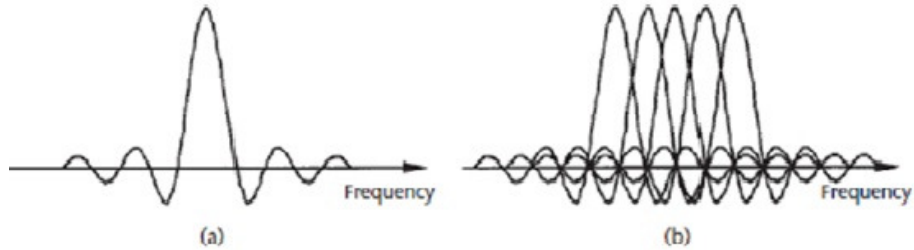

Figure 1. Spectra of (a) An OFDM Sub-channel and (b) An OFDM Signal [4]

In multicarrier transmission, bandwidth divided in many non-overlapping subcarriers but not necessary that all subcarriers are orthogonal to each other as shown in figure 1 (a). In OFDM the sub-channels overlap each other to a certain extent as can be seen in figure 1 (b), which leads to a proficient use of the total bandwidth. The information sequence is mapped into symbols, which are distributed and sent over the N sub-channels, one symbol per channel. To permit dense packing and still ensure that a minimum interference between the sub-channels is encountered, the carrier frequencies must be chosen carefully according to their orthogonal properties. By using orthogonal carriers, frequency domain can be viewed so as the frequency space between two sub-carriers is given by the distance to their first spectral null [4].

### 2.1. Mathematical Explanation of OFDM Signals

Consider, a data stream with rate R bps where bits are mapped to some constellation points using a digital modulation (QPSK or QAM). Let, N of these constellation points be stored for an interval of $T_s = N/R$, referred to as the OFDM symbol interval. A serial-to-parallel converter is used to achieve this. Now, each one of the N constellation points is used to modulate one of the subcarriers. Then, all modulated subcarriers are transmitted simultaneously over the symbol interval $T_s$ to get the proper OFDM signal [2]. The OFDM signal $x(t)$ can be expressed as,

$$x(t) = \sum_{k=0}^{N-1} a_k \exp(j2\pi(f_c + k\Delta f)t)$$

$$= \exp(j2\pi f_c t) \sum_{k=0}^{N-1} a_k \exp(j2\pi k\Delta f t)$$

$$= \exp(j2\pi f_c t) \, a(t) \qquad\qquad (1)$$

Where, $a_k$, $0 \le k \le N-1$, are complex-valued constellation points representing data and $f_k = f_c + k\Delta f$, $0 \le k \le N-1$, is the k$^{th}$ subcarrier, with $f_c$ being the lowest subcarrier frequency. $\Delta f$ is the frequency spacing between adjacent subcarriers, chosen to be $1/T_s$ to ensure that the subscribers are orthogonal. However, OFDM output symbols typically have large dynamic envelope range due to the superposition process performed at the IFFT stage in the transmitter.

## 3. SYNOPSIS OF PAPR

PAPR is extensively used to evaluate this variation of the output envelope. It is also an important factor in the design of both high power amplifier (PA) and digital-to-analog (D/A) converter, for generating error-free (minimum errors) transmitted OFDM symbols. As, there are large number of independently modulated sub-carriers are existed in an OFDM system, the peak value of the system can be very large as compared to the average value of the whole system. Coherent addition of N signals of same phase produces a large peak which is N times of the average signal. So, the ratio of peak power to average power is known as PAPR.

$$PAPR = \frac{Peak\ Power}{Average\ Power}$$

The PAPR of the transmitted signal is defined as [5],

$$PAPR = \frac{\displaystyle\max_{0 \le t \le NT} |x(t)|^2}{\dfrac{1}{NT} \displaystyle\int_0^{NT} |x(t)|^2\ dt} \qquad (2)$$

## 4. AMPLITUDE CLIPPING AND FILTERING

Amplitude Clipping and Filtering is one of the easiest techniques which may be under taken for PAPR reduction for an OFDM system. A threshold value of the amplitude is fixed in this case to limit the peak envelope of the input signal [6].

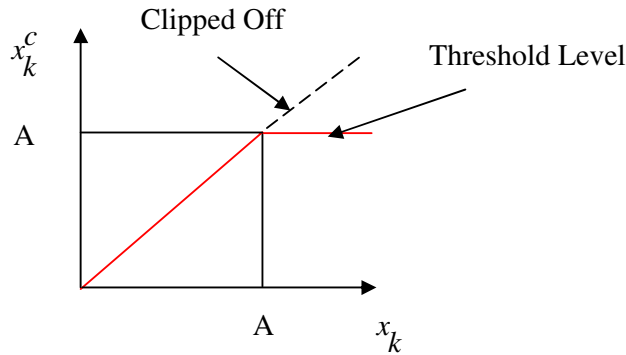

Figure 2.  Clipping Function

The clipping ratio (CR) is defined as,

$$CR = \frac{A}{\sigma} \qquad (3)$$

Where, A is the amplitude and $\sigma$ is the root mean squared value of the unclipped OFDM signal. The clipping function is performed in digital time domain, before the D/A conversion and the process is described by the following expression,

$$x_k^c = \begin{cases} x_k & |x_k| \leq A \\ Ae^{j\phi(x_k)} & |x_k| > A \end{cases} \qquad 0 \leq k \leq N-1 \qquad (4)$$

Where, $x_k^c$ is the clipped signal, $x_k$ is the transmitted signal, A is the amplitude and $\phi(x_k)$ is the phase of the transmitted signal, $x_k$.

## 4.1. Limitations of Amplitude Clipping and Filtering

> Clipping causes in-band signal distortion, resulting in BER performance degradation [7].
> Clipping also causes out-of-band radiation, which imposes out-of-band interference signals to adjacent channels. Although the out-of-band signals caused by clipping can be reduced by filtering, it may affect high-frequency components of in-band signal (aliasing) when the clipping is performed with the Nyquist sampling rate in the discrete-time domain. However, if clipping is performed for the sufficiently-oversampled OFDM signals (e.g., L $\geq$4) in the discrete-time domain before a low-pass filter (LPF) and the signal passes through a band-pass filter (BPF), the BER performance will be less degraded [7].
> Filtering the clipped signal can reduce out-of-band radiation at the cost of peak regrowth. The signal after filtering operation may exceed the clipping level specified for the clipping operation [8].

## 5. PROPOSED CLIPPING AND FILTERING METHOD

Indicating the second point of limitation [8] that is clipped signal passed through the BPF causes less BER degradation, we previously designed a scheme for clipping & filtering method where clipped signal would pass through a high pass filter (HPF) [9]. The proposed method is now shown in the figure 3. It shows a block diagram of a PAPR reduction scheme using clipping and filtering, where $L$ is the oversampling factor and $N$ is the number of subcarriers. The input of the IFFT block is the interpolated signal introducing $N(L-1)$ zeros in the middle of the original signal is expressed as,

$$X'[k] = \begin{cases} X[k], & for \quad 0 \leq k \leq \dfrac{N}{2} \quad and \quad NL - \dfrac{N}{2} < k < NL \\ 0 & Elsewhere \end{cases} \qquad (5)$$

In this system, the L-times oversampled discrete-time signal is generated as,

$$x'[m] = \frac{1}{\sqrt{LN}} \sum_{k=0}^{L N-1} X'[k] \cdot \exp\left(\frac{j2\pi n\Delta f\, k}{LN}\right), \qquad m = 0,1,...NL-1 \qquad (6)$$

and is then modulated with carrier frequency $f_c$ to yield a passband signal $x^p[m]$.

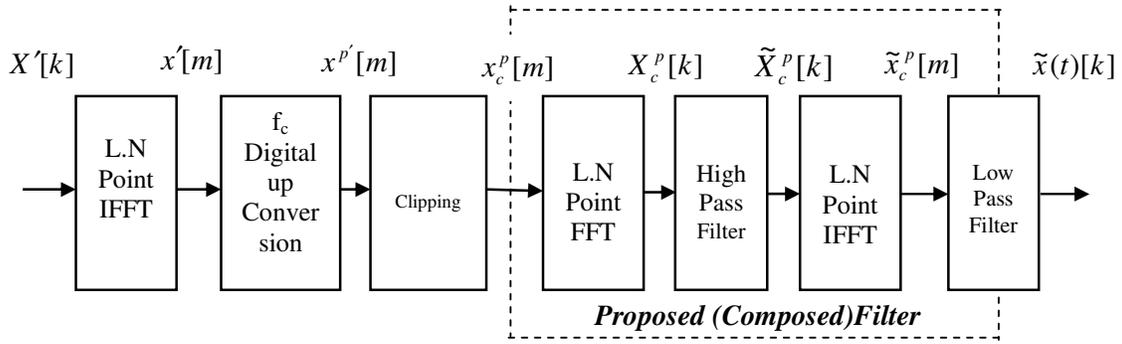

Figure 3.  Block Diagram of Proposed Clipping & Filtering Scheme.

Let, $x_c^p[m]$ denote the clipped version of $x^p[m]$ which is expressed as,

$$x_c^p[m] = \begin{cases} -A & x^p[m] \leq -A \\ x^p[m] & |x^p[m]| < A \\ A & x^p[m] \geq A \end{cases} \tag{7}$$

Where, *A* is the pre-specified clipping level.  After clipping, the signals are passed through the proposed filter (Composed Filter). The filter itself consists on a set of FFT-IFFT operations where filtering takes place in frequency domain after the FFT function. The FFT function transforms the clipped signal $x_c^p[m]$ to frequency domain yielding $X_c^p[k]$. The information components of $X_c^p[k]$ are passed to a high pass filter (HPF) producing $\tilde{X}_c^p[k]$. This filtered signal is passed to the unchanged condition of IFFT block and the out-of-band radiation that fell in the zeros is set back to zero. The IFFT block of the filter transforms the signal to time domain and thus obtain $\tilde{x}_c^p[m]$.

## 6. DESIGN AND SIMULATION PARAMETERS

In our previous research works, a linear-phase FIR filter using the Parks-McClellan algorithm was used in the composed filtering [9]. Existing method [7] uses the band pass filter. But, using this special type of high pass filter in the composed filter, significant improvement was observed in the case of PAPR reduction. The Parks-McClellan algorithm uses the Remez exchange algorithm and Chebyshev approximation theory to design filters with an optimal fit between the desired and actual frequency responses. The filters are optimal in the sense that the maximum error between the desired frequency response and the actual frequency response is minimized. The observations were actually based on only QPSK & QAM. In this simulation, using this filter, the effects of other higher order modulation techniques (8-PSK, 16-PSK, 32-PSK, 8-QAM, 16-QAM & 32-QAM) will be analyzed.

Table 1 shows the values of parameters used in the different modulation systems for analyzing the performance of clipping and filtering technique.

Table 1.  Parameters Used for Simulation of Clipping and Filtering.

| Parameters | Value |
|---|---|
| Bandwidth ( BW) | 1 MHz |
| Over sampling factor (L) | 8 |
| Sampling frequency, $f_s$ = BW*L | 8 MHz |
| Carrier frequency, $f_c$ | 2 MHz |
| No. of Subscribers (N) | 128 |
| CP / GI size | 32 |
| Clipping Ratio (CR) | 0.8, 1.0, 1.2, 1.4, 1.6 |
| Modulation Format | QPSK, 8-PSK, 16-PSK, 32-PSK, QAM, 8-QAM, 16-QAM & 32-QAM) |

## 6.1. Simulation Results for PAPR Reduction

In this first section, simulation is performed on our design for different higher order modulation techniques and analyzed their performances in case of reducing PAPR. Here, we want to monitor the effect of same number of symbol order (both for QPSK & QAM) step by step. It was analyzed QPSK with QAM previously. Now, other comparative analysis will be discussed in the next section.

### 6.1.1 Simulation Results:

In this section, PAPR distributions for different CR values are shown in the following figures. Clipped & filtered signal are shown in red colours.

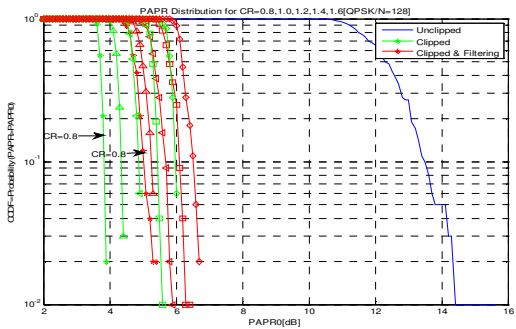
(a)

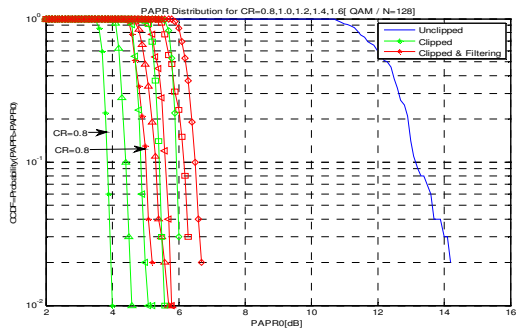
(b)

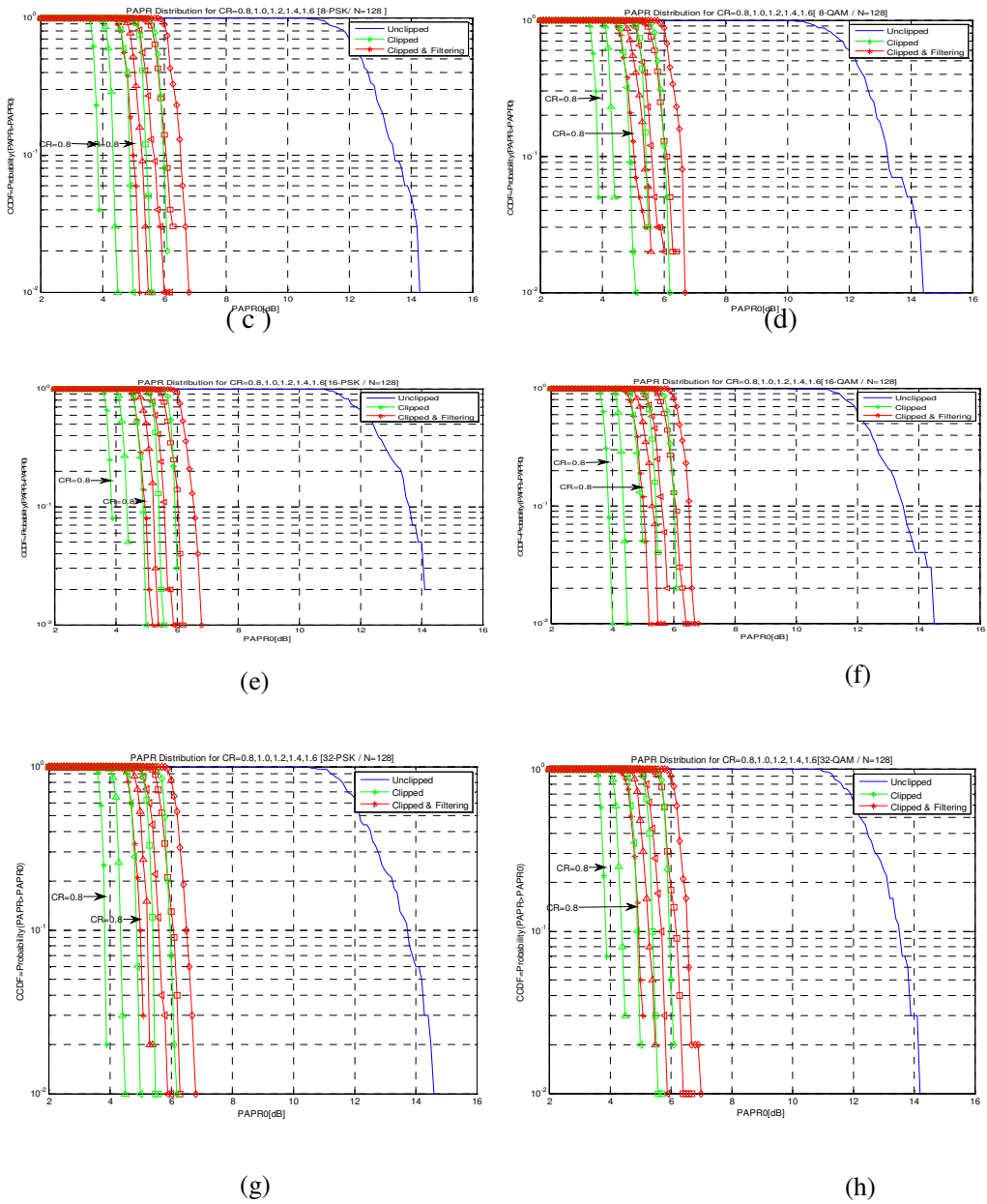

(c)

(d)

(e)

(f)

(g)

(h)

Figure 4.  PAPR distribution for CR=0.8, 1.0, 1.2, 1.4, 1.6;

| (a) QPSK and N=128; | (b) QAM and N=128 |
|---|---|
| (c) 8-PSK and N=128; | (d) 8- AM and N=128 |
| (e) 16-PSK and N=128; | (f) 16-QAM and N=128 |
| (g) 32-PSK and N=128; | (h) 32-QAM and N=128 |

In table 2, PAPR distribution for the above mentioned data are tabulated. The differences between same order modulations are also shown.

Table 2. PAPR Characteristics comparison of same symbol order modulation

| CR value | QPSK (dB) | QAM (dB) | Difference between QPSK & QAM (dB) | 8-PSK (dB) | 8-QAM (dB) | Difference between 8-PSK & 8-QAM (dB) |
|---|---|---|---|---|---|---|
| 0.8 | 5.11 | 4.97 | 0.14 | 5.001 | 5.038 | -0.037 |
| 1.0 | 5.18 | 5.25 | -0.07 | 5.281 | 5.37 | -0.089 |
| 1.2 | 5.65 | 5.67 | -0.02 | 5.601 | 5.618 | -0.017 |
| 1.4 | 6.04 | 6.09 | -0.05 | 6.061 | 6.101 | -0.04 |
| 1.6 | 6.51 | 6.51 | 0 | 6.570 | 6.569 | 0.001 |

| CR value | 16-PSK (dB) | 16-QAM (dB) | Difference between 16-PSK & 16-QAM (dB) | 32-PSK (dB) | 32-QAM (dB) | Difference between 32-QPSK & 32-QAM (dB) |
|---|---|---|---|---|---|---|
| 0.8 | 4.959 | 5.021 | -0.062 | 4.998 | 4.9 | 0.098 |
| 1.0 | 5.227 | 5.297 | -0.07 | 5.219 | 5.267 | -0.048 |
| 1.2 | 5.606 | 5.621 | -0.015 | 5.615 | 5.7 | -0.085 |
| 1.4 | 6.026 | 6.069 | -0.043 | 6.064 | 6.174 | -0.11 |
| 1.6 | 6.552 | 6.552 | 0 | 6.499 | 6.498 | 0.001 |

Performance Analysis:

Firstly, for the same number of subscribers (N=128) & low CR=0.8, QAM provides less PAPR than QPSK. But, at the moderate CR value (1.0, 1.2, 1.4), QPSK results less PAPR than QAM. At the high CR value (1.6), there is no difference between using QAM & QPSK. So, for lower CR (More Amount of Clipping), QAM is more suitable than QPSK for this design.

Secondly, it is examined that for the symbol order (8), 8-PSK shows the less PAPR than 8-QAM up to the CR value (1.4). But, at the higher CR value (Less Amount of Clipping), 8-QAM provides the better results.

Thirdly, it is found that for the symbol order (16), 16-PSK shows the less PAPR than 16-QAM up to the CR value (1.4). But, at the higher CR value (Less Amount of Clipping), both formats provide the same results

Lastly, it is observed that for the higher symbol order (32), 32-PSK shows the less PAPR than 32-QAM up to the CR value (1.4). But, at the higher CR value (Less Amount of Clipping), 32-QAM provides the better results.

So, analyzing the simulated results by this design, it is clearly monitored that in case of higher CR value (Less Amount of Clipping), QAM is more appropriate than PSK. On the other hand, PSK is better suited than QAM in case of low CR value (More Amount of Clipping).

## 6.2. Simulation Results for BER Performance

The clipped & filtered signal is passed through the AWGN channel and BER are measured for different modulation techniques. It is shown from these figures that the BER performance becomes worse as the CR decreases. That means, for low value of CR, (More amount of clipping), the BER is more.

### 6.2.1 Simulation Results:

In this section, BER Performance for different CR values is shown in the following figures.

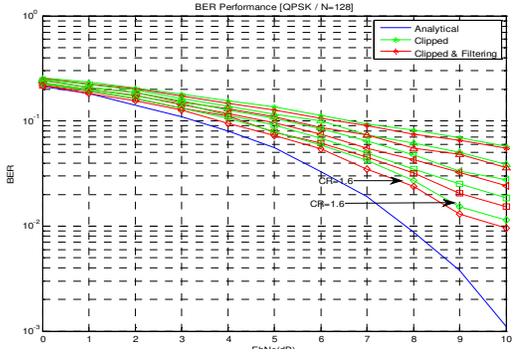

(a)

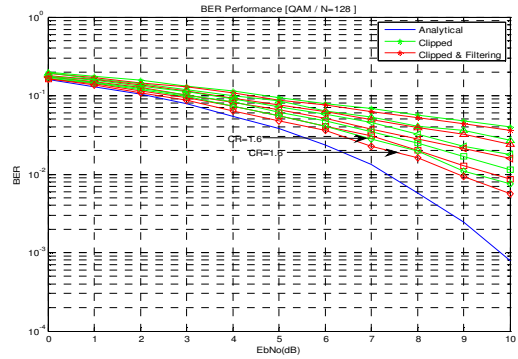

(b)

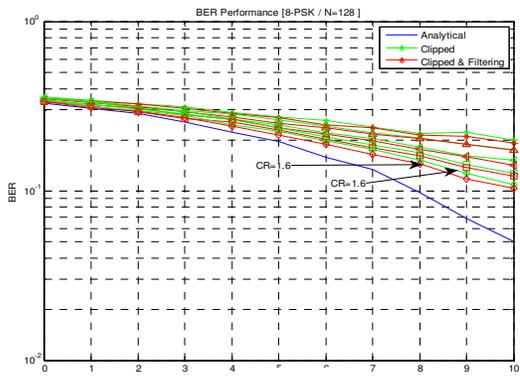

(c)

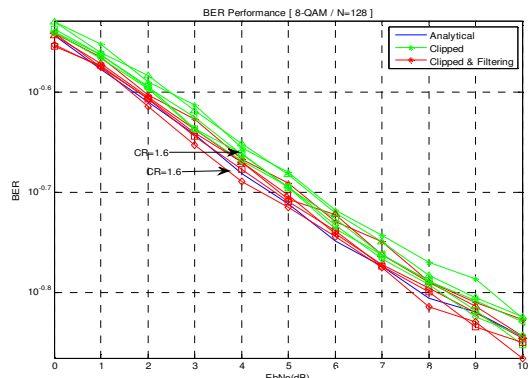

(d)

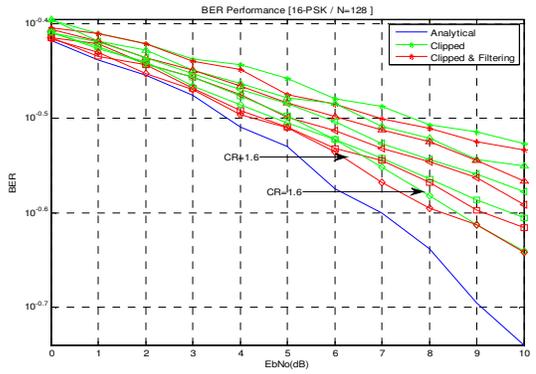

(e)

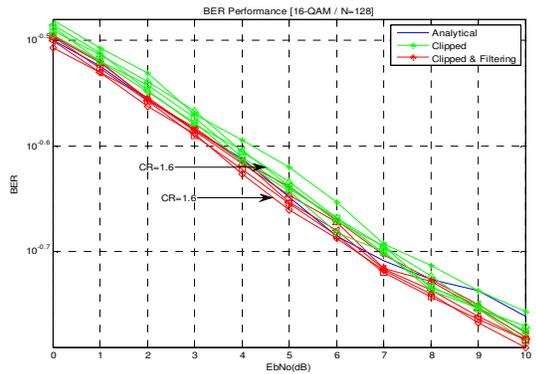

(f)

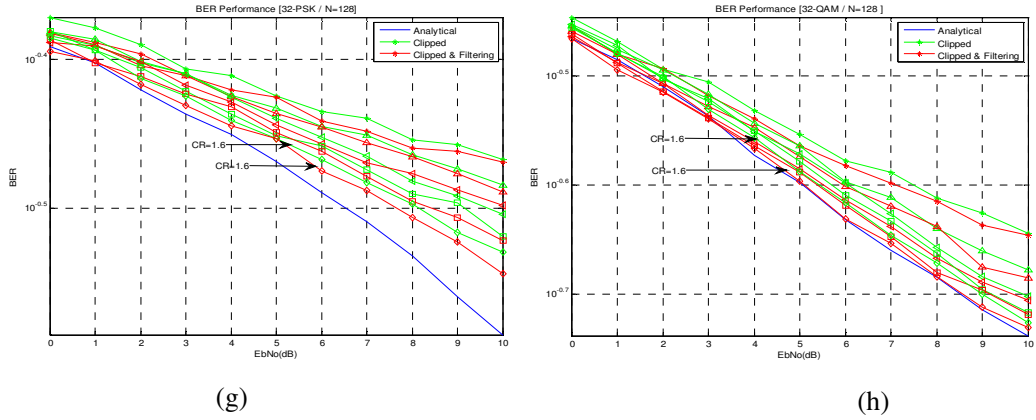

(g)                                                    (h)

Figure 5.  BER performance for CR=0.8, 1.0, 1.2, 1.4, 1.6;

(a) QPSK and N=128;          (b) QAM and N=128
(c) 8-PSK and N=128;         (d) 8- AM and N=128
(e) 16-PSK and N=128;        (f) 16-QAM and N=128
(g) 32-PSK and N=128;        (h) 32-QAM and N=128

Table 3. BER Performance comparison of same symbol order modulation

| CR value | QPSK | QAM | Difference between QPSK & QAM | 8-PSK | 8-QAM | Difference between 8-PSK & 8-QAM |
|---|---|---|---|---|---|---|
| 0.8 | 0.0752 | 0.07602 | -0.00082 | 0.2445 | 0.1896 | 0.0549 |
| 1.0 | 0.0616 | 0.06256 | -0.00096 | 0.2356 | 0.1865 | 0.0491 |
| 1.2 | 0.0492 | 0.05091 | -0.00171 | 0.2166 | 0.1827 | 0.0339 |
| 1.4 | 0.04025 | 0.04089 | -0.00064 | 0.2007 | 0.1815 | 0.0192 |
| 1.6 | 0.0339 | 0.03642 | -0.00252 | 0.1876 | 0.1803 | 0.0073 |

| CR value | 16-PSK | 16-QAM | Difference between 16-PSK & 16-QAM | 32-PSK | 32-QAM | Difference between 32-QPSK & 32-QAM |
|---|---|---|---|---|---|---|
| 0.8 | 0.3279 | 0.2137 | 0.1142 | 0.3617 | 0.2618 | 0.0999 |
| 1.0 | 0.3176 | 0.2129 | 0.1047 | 0.3583 | 0.2506 | 0.1077 |
| 1.2 | 0.3071 | 0.2088 | 0.0983 | 0.3482 | 0.2436 | 0.1046 |
| 1.4 | 0.2939 | 0.2067 | 0.0872 | 0.3452 | 0.2408 | 0.1044 |
| 1.6 | 0.2914 | 0.2053 | 0.0861 | 0.3349 | 0.2339 | 0.1010 |

Performance Analysis:

It is observed from the table 3 that, for all CR values, QAM results more BER than QPSK. But interestingly, in case of higher order modulation, M-QAM provides less BER than M-PSK (M=8, 16 and 32). As stated earlier, that for low CR means more amount of clipping that consequences more amount of BER, so, it is also monitored that for all cases of modulation.

As, PAPR reduction using amplitude clipping & filtering is a distortion method, so, there is a need for system trade-off. Here, we reduce PAPR with a little sacrifice of BER.

From data, it is also analyzed that, in case of QPSK & QAM, for almost 3% reduction of PAPR causes maximum 4% increment of BER i.e:~ that is acceptable. But, for this design, using higher order modulation, almost 3% PAPR reduction, it causes more than 15% BER i.e:~ that is too much.

Another viewpoint is the bit error rate per dB (BER/dB) shows that for the differences between same order modulations (M-PSK & M-QAM) gradually increases as M increases in case of a particular CR.

# 6. CONCLUSION

In this paper, a comparative performance is analyzed using higher order modulation techniques. It is observed from the simulated result that using this design, in case of higher CR value (Less Amount of Clipping), QAM is more appropriate than PSK. On the other hand, PSK is better suited than QAM in case of low CR value (More Amount of Clipping). As QPSK provides less PAPR than QAM, so, it causes high BER compare to QAM. The rational amount of BER is quite more than rational amount of PAPR reduction in case of all higher order modulations. So, in this design, lower order modulation (M=4) is better than higher order modulations (M=8, 16 and 32). In the present simulation study, ideal channel characteristics have been considered. In order to estimate the OFDM system performance in real world, multipath fading will be the next concern. The increase number of subscribers (N) & other parameters can be another assumption for further study.